\documentstyle[prl,aps,multicol,epsf]{revtex}

\newcommand{\hide}[1]{}
\newcommand{\gtopl}{\mbox{\hbox{ \lower-.6ex\hbox{$>$}\kern-.8em
\lower.5ex\hbox{$<$}\kern+.35em}}}

\title{Absence of Reentrance in the Two-Dimensional XY-Model with
Random Phase Shift}

\author{Thomas Nattermann$^{(a)}$, Stefan Scheidl$^{(a)}$,
Sergey E. Korshunov$^{(b)}$, and Mai Suan Li$^{(c)}$}

\address{$^{(a)}$ Institut f\"ur Theoretische Physik,
Universit\"at zu K\"oln,
D-50937 K\"oln,
Germany\\
$^{(b)}$ Landau Institute for Theoretical Physics, Kosygina 2,
117940 Moscow, Russia\\
$^{(c)}$ Institute of Physics,
02-668 Warsaw,
Poland}

\date{February 1995}
\begin{document}
\maketitle
\begin{abstract}
We show, that the 2D XY-model with random phase shifts exhibits for
low temperature and small disorder a phase with quasi-long-range
order, and that the transition to the disordered phase is {\it not}
reentrant. These results are obtained by heuristic arguments, an
analytical renormalization group calculation, and a numerical
Migdal-Kadanoff renormalization group treatment. Previous predictions
of reentrance are found to fail due to an overestimation of the vortex
pair density as a consequence of independent dipole approximations. At
positions, where vortex pairs are energetically favored by disorder,
their statistics becomes effectively fermionic. The results may have
implications for a large number of related models.
\end{abstract}
\begin{multicols}{2}
\narrowtext

\newpage

We reconsider in this paper the 2-dimensional XY-model
\begin{equation}
{\cal H} = - J\sum_{<i,j>} \cos (\phi_i - \phi_j - A_{ij})
\end{equation}
with quenched random phase shifts $A_{ij}$ on the bonds, where $i,j$
run over the sites of a square lattice. For simplicity we assume,
that the $A_{ij}$ on different bonds are uncorrelated and gaussian
distributed with mean zero and variance ${\sigma}$.

Model (1) describes e.g. 2-dimensional XY-magnets with random
Dzyaloshinskii-Moriya interaction \cite{RSN}. Other realizations are given by
Josephson-junction arrays with positional disorder \cite{jja} and model
vortex glasses \cite{mvg}. In particular, in the case of the so-called gauge
glass model, one assumes $A_{ij}$ to be uniformly distributed
between $0$ and $2\pi$. We expect, that our model with gaussian
disorder is equivalent to the gauge glass model when $\sigma \to \infty$.

For vanishing $A_{ij}$ model (1) undergoes a Kosterlitz-Thouless (KT)
transition, at which the spin-spin correlation exponent $\eta$ jumps
from 1/4 to zero \cite{KT}.

Weak disorder, $\sigma \ll 1$, should not change much this picture. In
the spin wave approximation one obtains $\eta$ = $\frac {1}{2\pi} (T/J
+\sigma)$, which remains now finite even at $T = 0$.  The features of
the KT-transition are essentially preserved, but the transition is
shifted to lower temperatures and the jump of $\eta $ at the
transition is diminished \cite{RSN}.  The actual transition temperature
$T_c(\sigma) \le T_+(\sigma)$ depends on the bare value for the vortex
core energy $E_c$, here $T_{\pm }=\frac{\pi}{4}J[1 {\pm
}{(1-8\sigma/\pi)}^{1/2}]$.  In the limit $E_c \to \infty $, $
T_c=T_+$.

Strong disorder will suppress the quasi-long-range order of the KT
phase \cite{mvg}. In particular, if $Q = \frac{1}{2\pi}\sum_{<plaq>}A_{ij}$
is of the order one, vortices are generated even at zero
temperature. Here $\sum_{<plaq>}$ denotes the sum over the four bonds
of an elementary plaquette.

Rubinstein, Shraiman and Nelson (RSN) \cite{RSN} extended the Coulomb gas
description of the KT-transition \cite{KT} to the presence of randomly
frozen dipoles arising from the random phase shifts. Surprisingly,
they found a second (reentrant) transition at $T_{re}(\sigma)$ $(\le
T_-(\sigma))$ to a disordered phase at low temperatures (see
Fig. 1). $T_{re}(\sigma)$ bends towards higher temperatures for
increasing disorder. The two lines $T_{\pm} $ merge at $\sigma _c =
\pi /8$. For $\sigma>\sigma _c$ there is no ordered phase.
The precise value of $T_{re}(\sigma)$ depends again on $E_c$. Similar
results were obtained in Ref. [2].

Korshunov [5] has argued, that the intermediate phase in the range
$T_{re}(\sigma) < T < T_c(\sigma)$ with quasi-long-range order is
probably {\it not stable}, if in addition to the screening of Coulomb
charges by neutral pairs of charges, considered in [1], screening by
larger complexes of charges in different replicas are taken into
account.

Experiments [6] as well as Monte Carlo studies [7] indicate no
reentrance. Also, Ozeki and Nishimori [8] have shown for a general
class of random spin systems, which include (1), that the phase
boundary between the KT- and the paramagnetic phase is parallel to the
temperature axis for {\it low T}. Thus they exclude a reentrant
transition, provided the intermediate KT phase exists. However, they
cannot rule out the possibility, that the KT-phase {\it disappears}
completely, as suggested in [5].

We will argue below, that the reentrant transition is indeed an
artefact of the calculation scheme used in [1], [2] and that the
KT-phase is stable at low temperatures with $T_c(\sigma) \to 0$ for
$\sigma \to \pi /8$ (see Fig.1). Since the renormalization group (RG)
flow equations (4) (see below) derived in [1], which give rise to the
reentrant behavior, appear as a subset of the more general RG
equations for XY-systems with additonal symmetry-breaking [9] or
random fields [10], as well as for solid films with quenched random
impurities [11], also these systems have to be reconsidered, which we
will postpone to forthcoming publications.

For the further discussion it is useful to decompose the Hamiltonian
(1) into a spin-wave part ${\cal H}_{sw}$ and a vortex part ${\cal
H}_v$. Since the phase transition is governed by ${\cal H}_v$, we will
omit ${\cal H}_{sw}$ completely. In the continuum description, the
vortex part can be rewritten in the form [1] (for simplicity, we set
the lattice constant equal to unity)
\begin{eqnarray}
{\cal H}_v &=& - J \pi \sum _{i} m_i \{\sum_{j \not= i}m_j \ln |{\bf r}_i -
{\bf r}_j | +\nonumber \\
&+& 2\int d^2r Q({\bf r}) \ln |{\bf r}-{\bf r}_i|- \frac{E_c}{J\pi}m_i\} .
\end{eqnarray}

The integer vortex charges $m_i$ satisfy $\sum _i m_i =0$. $Q({\bf
r})$ is a quenched random charge field, which is related to the phase
shift ${\bf A} ({\bf r})$ by $2\pi Q(r) = -\partial_xA_y+
\partial _yA_x$. Here we made the replacement $A_{ij} \rightarrow {\bf A}
({\bf r})$ by going over to the continuum description. Since
\begin{equation}
[{\bf A}]_d = 0, \qquad [A_\alpha({\bf r})A_\beta({\bf r'})]_d = \sigma
\delta _{\alpha \beta}\ ({\bf r}-{\bf r'}),
\end{equation}
where $[...]_d$ denotes the disorder average, the random charges are
{\it anticorrelated}.

The main result of the work of RSN [1] are the RG-flow equations
(4a-c) (see also [2], [9], [10]), which describe the change of $J,
\sigma $ and the vortex number density $y$ after eliminating vortex
degrees of freedom up to a length scale $e^l$
\begin{mathletters}
\begin{eqnarray}
\frac{dJ}{dl}&=& -4{\pi}^3 \frac{J^2}{T} y^2\label{4a}\\
\frac{dy}{dl}&=&(2-\pi \frac{J}{T} +\pi\frac{J^2}{T^2} \sigma)y\label{4b}\\
\frac{d\sigma}{dl}&=&0.\label{4c}
\end{eqnarray}
\end{mathletters}
Here we use the convention, that only the exchange constant $J$ is
renormalized and the temperature plays merely the role of an
unrenormalized parameter.  For $\sigma \equiv 0$ the equations (4a),
(4b) behave well defined for $T\rightarrow 0$.  This becomes more
clear, if we rewrite the vortex fugacity as $y=e^{-F_c/T}$, where
$F_c$ is the (core) free energy of a single vortex on the scale
$e^l$. Then (4b) takes the form ${dF_c}/{dl}=(\pi J - 2T - \pi
\frac{J^2}{T}\sigma )=2(T_+-T)(T-T_-)/T$.

For $\sigma > 0$, the last term on the r.h.s. of (4b) blows up at low
$T$, leading to the reentrance transition mentioned above. Whereas for
high temperatures the $1/T$ coefficient of the $\sigma$ term is
plausible, since thermal fluctuations wipe out the random potential,
we do not see a reason that this effect could lead to an {\it
unlimited} growth of the effective disorder strength at very low
temperatures. Clearly, (4b) cannot be valid at zero
temperature. Contrary to RSN [1], we argue, that the equations (4a),
(4b) are valid only for sufficiently high temperatures $T\ge
T^*(\sigma)>T_-(\sigma)$.

An indication for $T^*$ follows from the flow of the vortex entropy
$S_c = -{\partial F_c}/{\partial T}$, $ {\partial S_c}/{\partial l}
=2-\pi\frac{J^2}{T^2}\sigma+ \pi(-\frac {\partial J}{\partial
T})(1-2\sigma\frac {J}{T})$.  Since ${\partial J}/{\partial T} \le 0$,
the entropy is {\it reduced} for $T<T^*=2J\sigma$, $\sigma\le \pi/8$,
if one goes over to larger length scales. This leads finally to a
negative entropy, which we consider as an artefact of the calculation
[1] (see also [2], [9], [10], [11]).
%Similarly, the vortex
%core energy $E_c = F_c + TS_c$ flows to lower values for $T<2J\sigma $.
The vanishing of the entropy in disorderd systems usually signals a
freezing of the system by approaching $T^*$ from high temperatures
[12]. Similarly, the flow of the vortex energy $E_c=F_c+T S_c$, $
{\partial E_c}/{\partial l} =(1-2\sigma\frac {J}{T}) \pi (
J-T\frac{\partial J}{\partial T})$ leads for $T<T^*$ eventually to
negative values of the core energy. Inevitably, this favours multiple
occupancy of vortex positions. However, the resulting vortices of
higher vorticity $|m|>1$ appear even in the presence of disorder much
less likely than those with $|m|=1$: since their energy cost scales as
$m^2$ whereas their energy gain scales only as $m$. This effective
repulsion of vortices leads for $T<T^*$ to a much smaller vortex
density than in the RSN-theory [1], which neglects completely the
interaction between vortex dipoles.

For $T<T^*$ we expect the physics to be different from that described
by Eqs. (4). Since $T^*(\sigma)$ intersects the RNS phase boundary at
$\sigma = \pi /8$ where $T_+=T_-=J\pi/4$, the whole $(T,\sigma)$-range
in which reentrance was observed belongs to the freezing region, which
has to be reconsidered.

To find the correct behaviour at low temperatures, we consider first
the system at $T=0$. A simple estimate shows, that then vortices will
not be relevant if the disorder is weak. Indeed, the elastic energy of
an isolated vortex of charge $ \pm m$ in a system of radius $R$ is
$m^2 \pi J \ln R$, which has to be compared with the possible energy gain
$E_r$ from the interaction of the vortex with the disorder. If we
rewrite the second term in (2) as $\sum_i m_iV({\bf r}_i)$, we find
$[V^2({\bf r}_i)] _d \simeq 2\pi J^2 \sigma \ln R$.  Hence the {\it
typical} energy gain is $-J (m^2 2\pi \sigma \ln R)^{1/2}$.

In order to find the {\it maximal} energy gain, we have to estimate the
number $n(R)$ of vortex positions ${\bf r}_i$ in which the energies
$V({\bf r}_i)$ are essentially uncorrelated. Two vortex positions
${\bf r}_i, {\bf r}_j$ have independent energies if $[V^2({\bf
r}_i)]_d \gg [V({\bf r}_i)V({\bf r}_j)]_d$ , a condition which can be
rewritten with
\begin{equation}
[(V({\bf r}_i)-V({\bf r}_j))^2]_d = 4\pi \sigma J^2\ln |{\bf r}_i-{\bf r}_j|
\equiv \Delta ^2({\bf r}_i - {\bf r}_j)
\end{equation}
as $\ln|{\bf r}_i-{\bf r}_j| \approx (1-\epsilon)\ln R$ with $\epsilon
\ll 1$. Thus $n(R) \approx R^{2\epsilon}$ and the maximal energy gain
from exploiting the tail of the gaussian distribution for $V({\bf
r}_i)$ is $E_r\approx -2J(m^2 \pi \epsilon \sigma)^{1/2}\ln R$. The total
vortex free energy at $T=0$ is therefore
\begin{equation}
F_c \approx J\pi (m^2-2(m^2 \epsilon\sigma/\pi)^{1/2})\ln R
\end{equation}
and hence vortices should be irrelevant for weak disorder $\sigma \ll 1$.

In studying the behavior for $T=0$ but larger $\sigma$ we have to take
into account the screening of the vortex and quenched random charges
by other vortex pairs. This can be done most easily by using the
dielectric formalism. Here we follow the treatment of Halperin [13]
who showed, that screening by vortex pairs with separation between $R$
and $R+dR$ changes the coupling constant $J(R)$ (which corresponds to
the inverse dielectric constant) as
\hide{
This can be done most easily by using the dielectric formalism. Here
we follow the treatment of Halperin [13] who showed, that the change
of the coupling constant $J(R)$, resulting from the screening by
vortex pairs with separation between $R$ and $R+dR$, can be written as
}%hide
\begin{equation}
J(R+dR) = J(R)-4\pi^2 \sum_{m>0} \alpha_m(R) J^2(R) 2 \pi R p_m(R) dR.
\end{equation}
Here $\alpha_m(R)$ is the polarizability and $p_m(R)$ is the
probability density of a pair with charge $+m$ at ${\bf r}_1$
and charge $-m$ at ${\bf r}_2$, $R=|{\bf r}_1- {\bf r}_2|$.

For the calculation of $p_m{(R)}$ and $\alpha_m(R)$ we use
the fact, that the interaction energy between a vortex pair and the
disorder is gaussian distributed with a width $\Delta (R)$. If the
density of pairs is sufficiently small, we may neglect the interaction
between pairs and write
\begin{eqnarray}
p_m{(R)}&=&\int^{-2|m|(\pi J\ln R+E_c)}_{-\infty}
\frac{dV}{\sqrt{2\pi} \Delta (R)}
e^{-\frac{V^2}{2\Delta ^2(R)}}\approx \nonumber \\
&\approx &\sqrt{\frac{\sigma}{2\pi^2 m^2 \ln R}}
 R^{-\frac{m^2 \pi}{2\sigma}},
\end{eqnarray}
where the $r.h.s.$ of (8) is valid only for $\sigma$, ${E_c}/{J} \ll
 \ln R$. Since $p_{|m|>1}(R) \ll p_1(R)=: p(R)$, we will
 neglect double occupancy of vortex positions.

The polarizability $\alpha(R)=\alpha_1(R)$ can be calculated in a
similar way and is found to be $\alpha (R) \approx R^2/T^*$ at large $R$.
 With $y^2=R^4 p(R)$ and $l=\ln R$
we get from (7) and (8) for $T=0$
\begin{mathletters}
\begin{eqnarray}
\frac{dJ}{dl}&=&- 4 \pi^3 \frac{J}{\sigma}y^2 \label{9a}\\
\frac{dy}{dl}&=&(2-\frac{\pi}{4\sigma})y \label{9b}\\
\frac{d\sigma}{dl}&=&0, \label{9c}
\end{eqnarray}
\end{mathletters}
where we again neglected terms of the order $\sigma/l$. These
are the flow equations, which replace (4) at {\it zero temperature}.
Within this approximation, the system undergoes a phase transition at
$\sigma _c=\pi/8$ from a KT to a disordered phase, which is in
qualitative agreement with our estimate (6). At $\sigma _c$ the
exponent $\eta$ shows a universal jump from $1/16$ to zero. For
$\sigma > \sigma_c$ the $y$ reaches a value of order magnitude unity
on the scale $R\approx \xi$ with
\begin{equation}
\xi\propto e^{1/b(1-\pi/{8\sigma})}.
\end{equation}
$b$ is a constant, which depends on the details of the system. For
$R>\xi$ our flow equations are no longer valid, since $y$ is no longer small.
We identify $\xi$ with the correlation length in the disordered phase.

We discuss now the properties of the system at low but {\it finite}
temperatures. The $T$-correction to our free energy estimate (6) are
of the order $-2T\ln R$ (or smaller) and hence will not allow a
reentrance transition. A more efficient way for thermal fluctuations
to influence the low-$T$ behaviour would be the generation of {\it
uncorrelated} frozen charges $Q(\bf r)$. However unlike to random
field systems, where {\it uncorrelated} random fields are indeed
generated from {\it anticorrelated} random fields [14], which destroy
the ordered phase in 2 dimension at all non-zero $T$, we do not see
such a mechanism here. The main difference consists in the existence
of a double degenerated ground state in the random field system at
$T=0$.

The physics at finite temperature can also be captured within the
dielectric formalism. Neglecting again the interaction between vortex
pairs at different positions, we calculate the {\it normalized}
probability for a pair with charges $\pm m$ as
\begin{equation}
p_m(R)=p_{-m}(R)=
\left[ \frac {e^{-E_m(R)/T}}{\sum_{m}e^{-E_m(R)/T}} \right]_d,
\label{def.p}
\end{equation}
where $E_m(R)=2 m^2 (E_c + \pi J \ln R) + m (V({\bf r}_1)- V({\bf
r}_2))$ denotes the pair energy. At large $R$ holds $p_{|m|>1}(R)
\approx 0$, since the elastic energy cost $\propto m^2$ will be
compensated with decreasing probability by an energy gain $\propto m$
due to disorder. We therefore drop occupancies $|m|>1$. Furthermore,
for a given configuration of disorder, one of the two energies $E_{\pm
1}(R)$ is always so large, that the corresponding weight factor
$e^{E_{\pm 1}(R)/T}$ can be neglected. After this approximation, the
probability for a single pair reads
\begin{equation}
p(R)= p_1(R)= \left[ \frac 1{1+e^{E_1(R)/T}} \right]_d.
\label{def.p1}
\end{equation}
Eq. (\ref{def.p}) thus effectively reduces to the disorder average of
the {\it Fermi distribution function}. In other words: vortex pairs of
vorticity one can be treated as {\it non-interacting fermions}. In the
limit $T=0$, where this distribution function becomes step-like,
Eq. (\ref{def.p1}) immediately reduces to the previous expression
(8). At finite temperature, the disorder average in Eq. (\ref{def.p1})
is performed by splitting the integral over the disorder distribution
into two contributions corresponding to $E_1(R) \gtopl 0$. To leading
order in $R$, we find $p(R) \sim R^{-\pi/2 \sigma}$ for $0 \leq T \leq
T^*=2J \sigma$, whereas $p(R) \sim R^{-2\pi J/T(1- \sigma J/T)}$ for
$T \geq T^*$. Plugging these results into the definition of $y$, we
obtain the flow equation (4b) in the whole range $T \geq T^*$, whereas
Eq. (9b) is valid in the whole range $0\leq T \leq T^*$.  Both
equations coincide at the boundary $T =T^*$.

We add a few remarks: As long as $E_1(R) \gg T$, the Fermi
distribution can be replaced by the Boltzmann distribution, as is
usually done in the treatment of the KT transition [4]. The disorder
average of the latter yields $p(R) \sim R^{-2\pi J/T(1- \sigma J/T)}$
and hence Eq. (4b) for {\it all} temperatures. However, for $T<T^*$
the condition $E_1(R) \gg T$ is no longer fulfilled for most of the
vortex positions (see also our remarks below Eqs. (4)) and hence this
approximation breaks down. Indeed, use of the Boltzmann distribution
at low temperatures would lead to $p(R) \gg 1$, and the interaction
between vortex pairs could no longer be neglected. It is therefore
important to calculate $p(R)$ from (\ref{def.p1}). An attempt to
improve upon the Boltzmann-approximation consists in expanding
(\ref{def.p1}) into a power series in $e^{-E_1(R)/T}$. The $n$-th
order term in the expansion yields a contribution $R^{-2\pi J/T(n-n^2
\sigma J/T)}$ to $p(R)$. The series is divergent, i.e. for large $R$
higher order terms are more important than lower order terms,
irrespective of temperature. These higher order terms generate
contributions to the flow equation $dy/dl$, which tend to blow up $y$
even faster. One might hence expect an instability of the ordered
phase, similarly to the observation of Korshunov [5]. In fact, the
above expansion and in particular the replacement of the Fermi- by the
Boltzmann distribution are disqualified {\it a posteriori}.

We conclude, that $dy(l)/dl <0$ for all $T<T_+$. The polarizability at
finite temperatures is given by $\alpha=R^2/(T+T^*)$ for $T<T^*$ and by
$\alpha=R^2/(2T)$ for $T>T^*$. Thus $dJ/dl<0$ holds for all $T<T_+$
which is sufficient to guarantee the absence of reentrant phase
topology. In the special case of $E_c \to \infty$ the phase boundary
is given by $T_+ (\sigma)$ for $T\ge J\pi/4$ and a horizontal line
$\sigma_c = \pi/8$ for smaller $T$, as shown bye bold lines in
Fig. 1. This is consistent with the prediction of Ozeki and Nishimori
[8] about the existence of a horizontal phase boundary. We expect the
critical behavior at $T_+(\sigma)$ as discussed in [1] to be
unchanged. At finite core energies, the actual transition temperature
will be renormalized to $T_c(\sigma)< T_+(\sigma)$. Its value $T_c(0)$
is given by the KT flow equations without disorder and lies only
slightly below $T_+(0)$ for large $E_c$. For small $\sigma$, the
critical RG trajectory flows completely in the domain of equation
(4b), where weak disorder induces weak additional screening. Therefore
$T_c(\sigma)$ will smoothly decrease with increasing $\sigma$. We
expect this function to end up in $T_c(\pi/8)=0$ monotonously, since
flow equations vary monotonously in parameter space.

Our conclusions about the absence of reentrance are confirmed also by
a discretized Migdal-Kadanoff renormalization group (MKRG) scheme [15]
for model (1), which we consider in the last part of this paper.  Our
technique has been shown to be similar to that of Jos\'e et
al. [16]. Their approach is based on studying Migdal-Kadanoff
recursion relations for the Fourier components of the (spatially
uniform) potential.

In the discretized scheme [15] instead of allowing $\phi$ to be a
continuous variable, we constrain it to take one of many discrete
values which are uniformly distributed between $0$ and $2\pi$.
Hamiltonian (1) is now defined for values of $\phi$ restricted to
$2\pi k/q$, where $k = 0, 1, 2, \ldots , (q-1)$ and $q$ is a number of
clock states. We define
\begin{equation}
J_{ij} (q,k) \; = \; J  \cos ( 2\pi k/q - A_{ij} ).
\label{eqn1}
\end{equation}
The recursion relations for $J_{ij}(q,k)$ may be found in [15].  For
the random 2D system, the numerical procedure is based on creating
first a pool of $N_p$ bonds, each decomposed into $q$ components
according to Eq. (\ref{eqn1}). One then picks $N_p$ random batches of
4 such bonds (the corresponding rescaling factor is equal to 2) from
the pool to generate a new pool of the coupling variables and the
whole procedure is iterated. We consider typically $N_p$=2000 and
$q$=100. The results depend on these parameters rather weakly.

It should be noted that Gingras and Sorensen [16] have tried to
construct the phase diagram of the 2D random Dzyaloshinskii-Moriya
model (this model is believed to be equivalent to (1) by the same
discretized MKRG approach. In order to locate the Kosterlitz-Thouless
phase, they study the scaling behavior of the absolute average height
of the potential, $\bar{h}$, which is defined as follows
\begin{equation}
\bar{h} \; = \; \langle | J_{ij}(q,0) - J_{ij}(q,q/4) | \rangle \; \; .
\label{eqn5}
\end{equation}
Due to errratic behavior of $\bar{h}$ they could not draw the phase
diagram. The reason here is that $\bar{h}$ representing only two
clock states cannot correctly describe the system with many clock states.

To obtain the phase diagram one can consider the scaling of the
maximal and minimal couplings for each effective bond or the scaling
of the average of absolute values of all $q$ couplings.  The scaling
properties of these three quantities has been found to be essentially
the same, so it is sufficient to focus on the maximum coupling
$J_{max}(q,k)$. The details of this approach can be found in
Ref. [15].

It should be noted that the discretized MKRG approach cannot
rigorously reproduce the quasi-long-range XY order in 2$D$ [15].  The
scale invariance of $J_{max}(q,k)$ in the KT-phase is merely
approximate in this approach.  In practice, the scale invariance of
$J_{max}(q,k)$ persists for about 20 iterations. Further iterations
lead to an eventual decrease of $J_{max}(q,k)$ at any nonzero
$T$. Having this caveat in mind, we can locate the boundary between
the paramagnetic and KT- phase (see Fig. 2). Thus the MKRG gives us
additional evidence that the reentrance is absent in model (1).

To conclude, in the present paper we have shown by a combination of
simple analytical arguments, a renormalization group calculation
and a Migdal-Kadanoff RG scheme at finite $T$, that the 2-dimensional
XY-model with random phase shifts does not exhibit a reentrant
transition.

{\bf Acknowledgments} T.N. acknowledges a stay at the ITP at Santa
Barbara where this work was begun.  It is a pleasure to thank
M. Kardar, H. Orland, L.H. Tang for discussion on this
problem. S.E.K. and M.S.L. acknowledge support from the SFB 341.

\begin{figure}[b]
\epsfxsize=\linewidth
%\epsfbox{fig1.eps}
\caption{$(\sigma,T)$ phase diagram of the model (1). $T_{\pm}(\sigma)$
are the {\it upper} bounds for the transition temperatures
$T_c(\sigma)$ and $T_{re}(\sigma)$ between the disordered and the KT
phase in the RSN-theory [1]. Note, that $T_-$-line lies completely in
the freezing region (hatched area). The true phase transition line
$T_c(\sigma)$ is denoted by the dashed line which is bounded by
$T_+(\sigma)$ and $\sigma = \pi/8$. The line $T_{re}$ is not shown
here.}
\end{figure}

\begin{figure}
\epsfxsize=\linewidth
%\epsfbox{fig2.eps}
\caption{$(\sigma,T)$ phase diagram obtained by the discretized
Migdal-Kadanoff RG scheme. PM and KT denote the paramagnetic and
Kosterlitz-Thouless phase respectively.  In the PM phase
$J_{max}(q,k)$ scales down monotonously whereas in the KT region it
reaches a fixed value at large scales. The critical values
$\sigma_c^{1/2} (T=0) \approx k_BT_c(\sigma =0)/J \approx 0.44$. One
can also demonstrate that the phase diagram of the random 2D
Dzyaloshinskii-Moriya model has the same topology.}
\end{figure}

\end{multicols}
\end{document}